\documentclass{ws-procs9x6}

\usepackage{graphicx}

\newcommand{\eVq}{\ensuremath{\text{eV}^2}}
\newcommand{\sgn}{\mathop{\rm sign}\nolimits}

\newenvironment{myitemize} {
  \begin{list}{--}
      {
	\setlength{\leftmargin} {4mm}
	\setlength{\parsep}     {0pt}
	\setlength{\itemsep}    {0mm}
	\setlength{\topsep}     {\itemsep}
	\setlength{\partopsep}  {0pt}
	\setlength{\parskip}    {0pt}
	}} {
  \end{list}}

\renewcommand{\cite}[1]{[\refcite{#1}]}

\begin{document}

\title{Status of Global Analysis of Neutrino Oscillation Data}

\author{M.C. GONZALEZ-GARCIA}

\address{Y.I.T.P., SUNY at Stony Brook, Stony Brook, NY 11794-3840, USA\\
  IFIC, Universitat de Val\`encia - C.S.I.C., Apt 22085, 46071
  Val\`encia, Spain }

\author{M. MALTONI}

\address{Y.I.T.P., SUNY at Stony Brook, Stony Brook, NY 11794-3840, USA}

%\preprint{YITP-SB-04-33}

\maketitle

\abstracts{%
  In this talk we discuss some details of the analysis of neutrino
  data and our present understanding of neutrino masses and mixing.
  This talk is based on Refs.~\cite{newthree,ourcpt,atmnp}.}

\section{Analysis of Solar and KamLAND}

\begin{figure}[t] \centering
    \includegraphics[width=37mm]{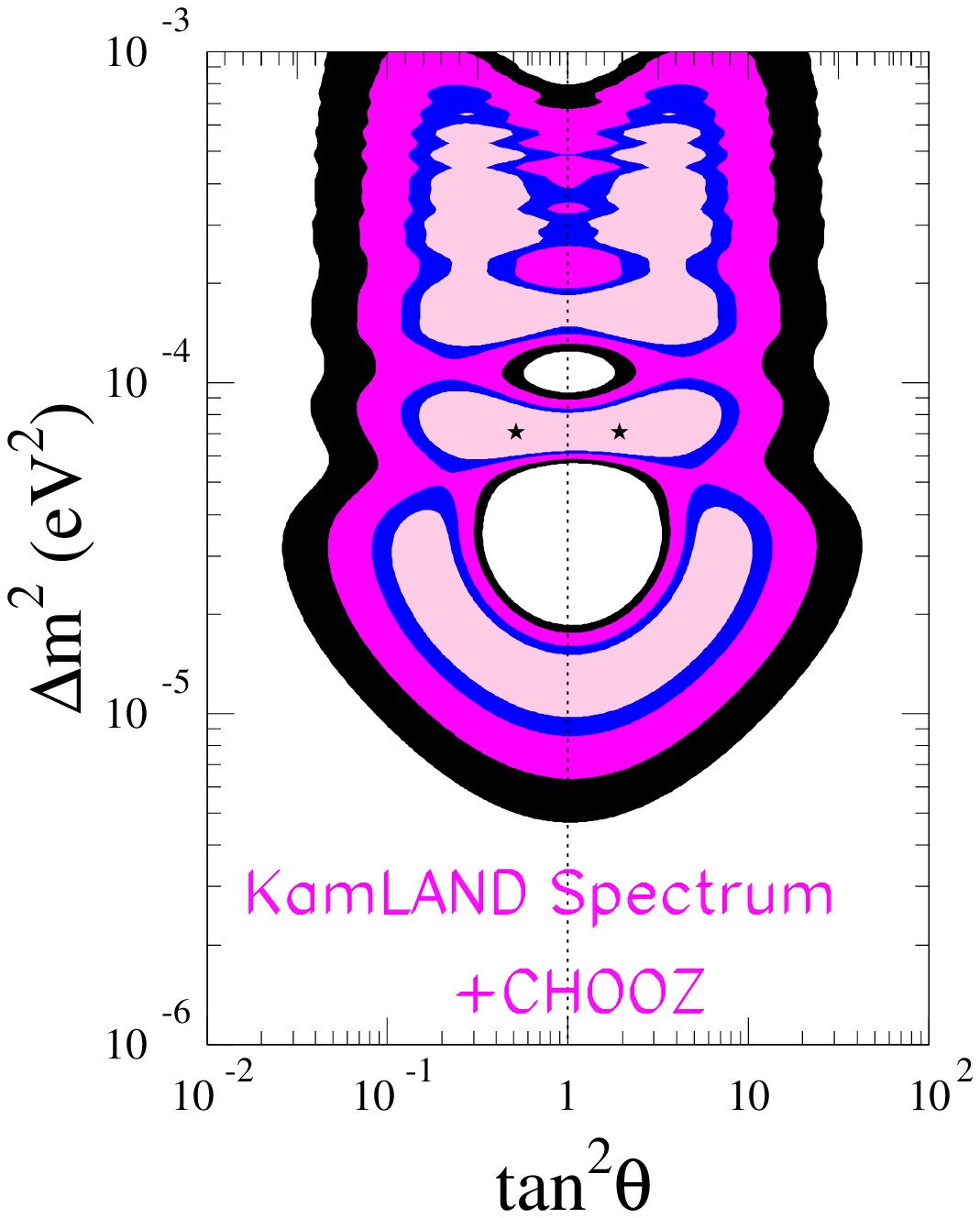}
    \hfill
    \raisebox{1.3mm}{\includegraphics[width=37mm]{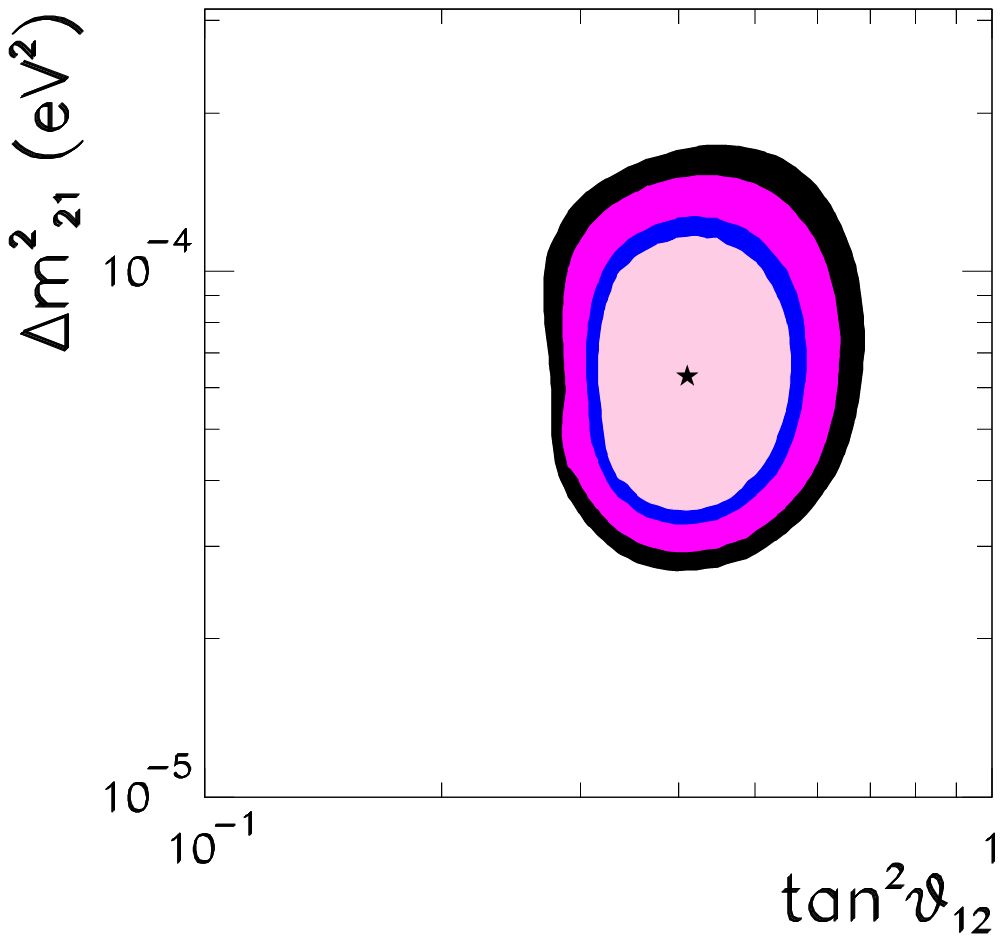}}
    \hfill
    \raisebox{0.8mm}{\includegraphics[width=37mm]{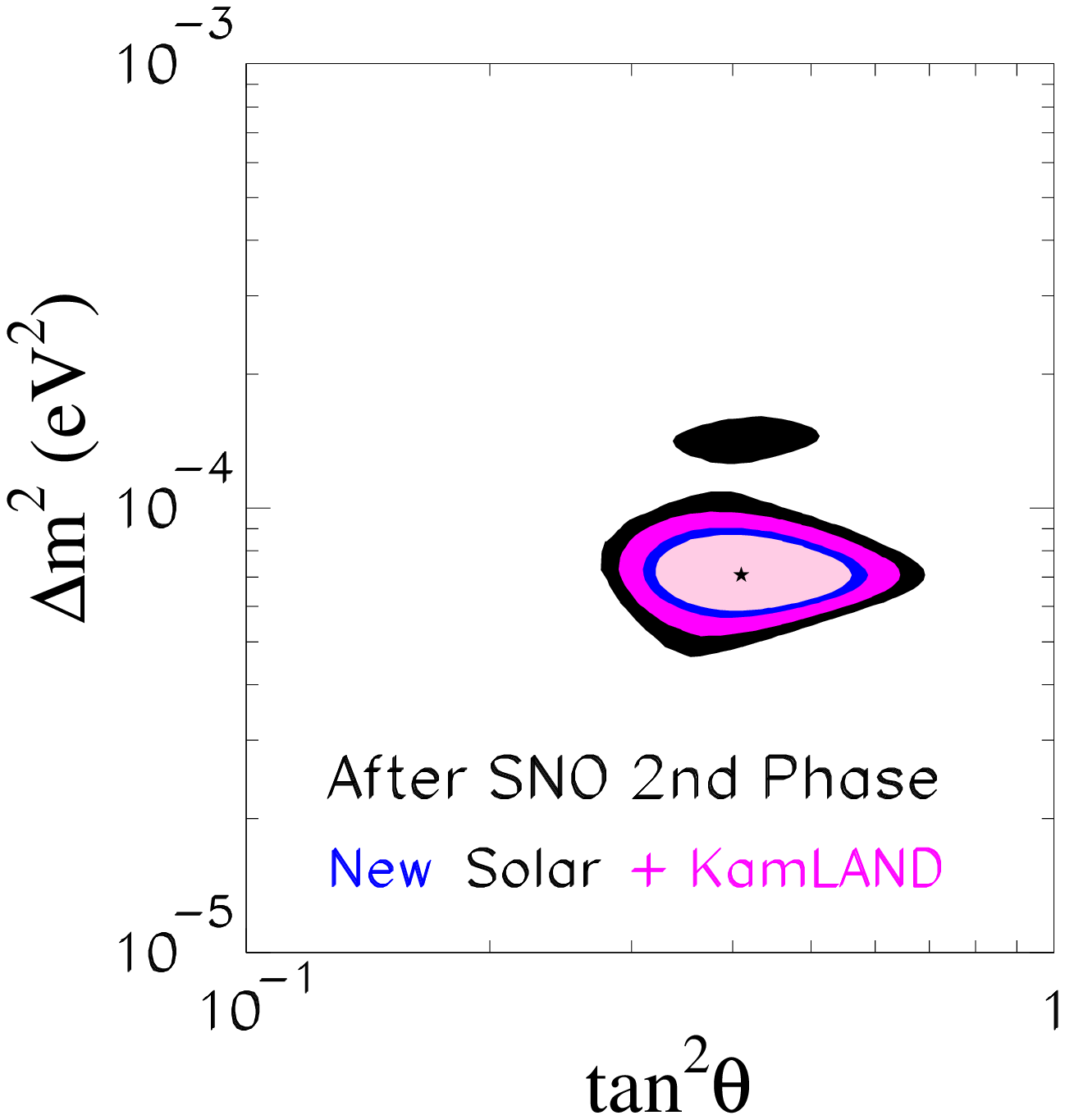}}
    \caption{\label{fig:kland}%
      Allowed regions for 2-$\nu$ oscillations of $\bar\nu_e$ in
      KamLAND and CHOOZ (left panel), of $\nu_e$ in the sun (central
      panel), and for the combination of KamLAND and solar data under
      the hypothesis of CPT conservation (right). The different
      contours correspond to the allowed regions at 90\%, 95\%, 99\%
      and $3\sigma$ CL.}
\end{figure}

In Fig.~\ref{fig:kland} we show the results from our latest analysis
of KamLAND~\cite{kland} reactor $\overline{\nu}_e$ disappearance data,
solar $\nu_e$ data~\cite{sksol,sno2} and the combined analysis under
the hypothesis of CPT symmetry. The main new ingredient in this
analysis with respect to the previous ones is the inclusion of the
first results from the SNO salt phase (SNOII) data~\cite{sno2}. We
have also taken into account the new gallium measurement which leads
to the new
average value $69.3 \pm 4.0$. The main changes as compared to the
pre-SNOII analysis are:
\begin{myitemize}
  \item in the analysis of solar data, only LMA is allowed at more than
    $3\sigma$;
  \item maximal mixing is rejected by the solar analysis at more
    than $5\sigma$;
  \item the combined analysis allows only the lowest LMA region at
    99.4\% CL;
  \item the new best-fit point is:
    \begin{equation}
	\Delta m^2=7.1\times 10^{-5}~\eVq \,, \qquad
	\tan^2\theta=0.41 \,, \qquad
	\frac{\Phi_{\rm^8B}}{\Phi_{\rm^8B, BP04}}=0.88 \,.
    \end{equation}
\end{myitemize}
These results are in agreement with those reported in the several
state-of-the-art analysis of solar and KamLAND data which exist in the
literature.  All these analysis share the same basic characteristics. 

In the analysis of KamLAND we used the following approximations:

\begin{myitemize}
  \item the antineutrino spectrum is parameterized~\cite{vogel} without
    detailed theoretical uncertainties;
  \item the yearly average reactor power is used.
\end{myitemize}
Presently the uncertainties of the KamLAND results are statistics
dominated so these effects do not make any difference in the extracted
allowed regions. The minor differences between the several 
phenomenological analysis in the literature are more likely to arise
from the use of different statistical functions in the analysis of
KamLAND data.

Before moving to atmospheric neutrinos, we wish to point out some
important features of the analysis of solar data:
\begin{myitemize}
  \item the SSM~\cite{carlos} provides detailed informations not only
    on the solar neutrino fluxes themselves, but also on their
    theoretical uncertainties and correlations due to variations of
    the SSM inputs;
  \item the spectral shape experimental uncertainty for $^8$B spectrum
    is properly taken into account;
  \item the energy dependence of the interaction cross sections
    and their uncertainties is also included;
  \item the interplay between the energy-dependent part of the
    theoretical and systematic uncertainties and the neutrino survival
    probability (which depends on the oscillation parameters) is
    properly taken into account.
\end{myitemize}

\section{Atmospheric Neutrinos}

\begin{figure}[t] \centering
    \includegraphics[width=0.9\textwidth]{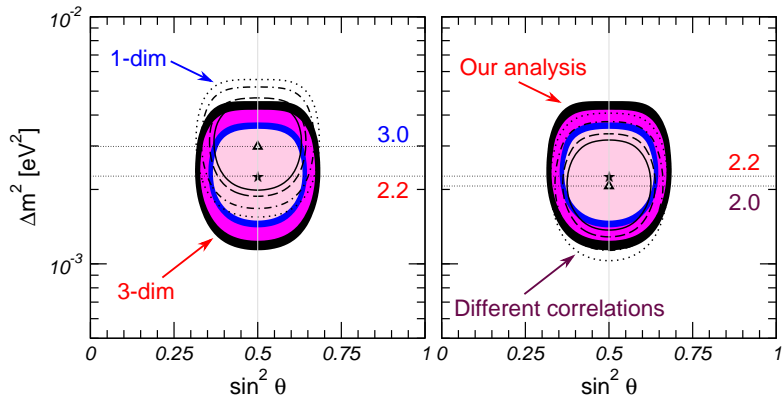}
    \caption{\label{fig:atm2fam} %
      \textit{Left:\/} allowed regions from the analysis of
      atmospheric data using the new (full regions labeled ``3-dim'')
      and old (empty curves labeled as ``1-dim'') SK1 data and
      atmospheric fluxes.  \textit{Right:\/} impact of a small change
      in the correlation between the theoretical uncertainties of the
      low-energy (sub-GeV) and high-energy (multi-GeV, and upgoing
      $\mu$) data samples. The different contours correspond to at
      90\%, 95\%, 99\% and $3\sigma$ CL.}
\end{figure}

In the left panel of Fig.~\ref{fig:atm2fam} we show the results of our
latest analysis of the atmospheric neutrino data, which included the
full data set of Super-Kamiokande phase I (SK1). As discussed in
Ref.~\cite{skatmos} the new elements in the Super-Kamiokande
analysis include:
\begin{myitemize}
  \item use of new three-dimensional fluxes from Honda~\cite{honda3d};
  \item improved interaction cross sections which agree better with
    the measurements performed with near detector in K2K~\cite{k2k};
  \item some improvements in the Monte-Carlo which lead to some
    changes in the actual values of the data points.
\end{myitemize}
We have included these elements in our calculations and we have also
improved our statistical analysis (see Ref.~\cite{atmnp} for
details).
Our results show good quantitative agreement with those of the Super-K
collaboration. In particular we find that after inclusion of the above
effects, the allowed region is shifted to lower $\Delta m^2$. The new
best-fit point is located at:
\begin{equation}
    \Delta m^2 = 2.2\times 10^{-3}~\eVq\,, \qquad
    \sin^2\theta=0.5 \,.
\end{equation}

\begin{figure}[t] \centering
    \includegraphics[width=3in]{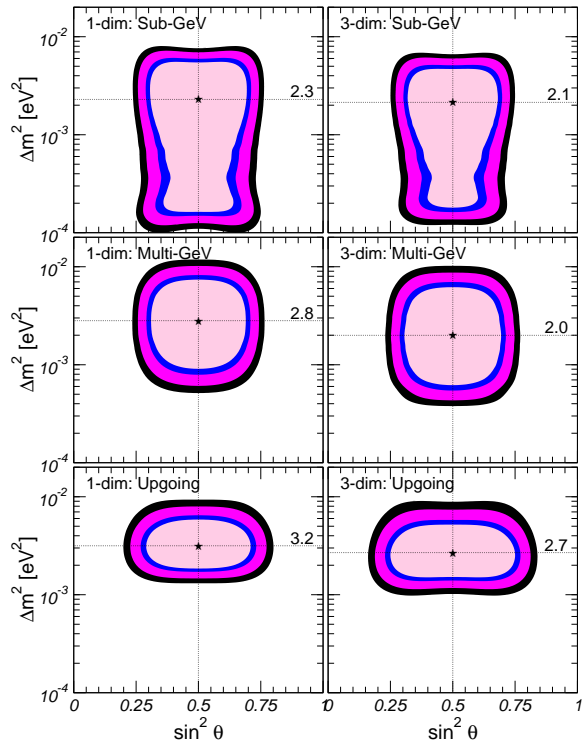}
    \caption{\label{fig:atmsets}%
      The left (right) panels show the allowed regions for the
      analysis of the different sets of atmospheric data using the new
      (old) SK1 analysis and atmospheric fluxes. The different
      contours correspond to the allowed regions at 90\%, 95\%, 99\%
      and $3\sigma$ CL.}
\end{figure}

At this point it is important to remark that, unlike for solar
neutrinos, the energy dependence of the theoretical uncertainties in
the atmospheric fluxes and in the interaction cross sections are not
so well determined in terms of a set of model inputs. To address this
issue, we have performed the same analysis assuming slightly different
correlations among the theoretical errors for the different data sets,
and we have found that the size of the final shift in $\Delta m^2$ of
the allowed region depends on these details (see right panel of
Fig.~\ref{fig:atm2fam}). The reason for this behavior is illustrated
in Fig.~\ref{fig:atmsets}, where we show the allowed regions obtained
with the new analysis as compared to the old one for the different
atmospheric data sets. As can be seen in the figure, the different
sets favor slightly different ranges of $\Delta m^2$, thus the
treatment of the energy dependence of the uncertainties becomes
relevant in outcome of the combined analysis. 

These results lead us to raise here a word of caution.  In all present
analysis of atmospheric data, two main sources of theoretical flux
uncertainties are included: an energy independent normalization error
and a ``tilt'' error which parametrizes the uncertainty in the
$E^{-\gamma}$ dependence of the flux. Some additional uncertainties in
the ratios of the samples at different energies are also allowed as
well as uncertainties in the zenith dependences.  However, we still
lack a well established range of theoretical flux uncertainties within
a given atmospheric flux calculation, in a similar fashion to what it
is provided for the solar neutrino fluxes by the SSM. In the absence
of these, we cannot be sure that we are accounting for the most
general characterization of the energy dependence of the atmospheric
neutrino flux uncertainties.

Given the large amount of data points provided by the Super-K
experiment, this is becoming an important issue in the atmospheric
neutrino analysis. There is a chance that the atmospheric fluxes may
be still too ``rigid'', even when allowed to change within the
presently considered uncertainties. As a consequence, we may be
over-constraining the oscillation parameters. 

\section{Three-Neutrino Oscillations}

The minimum joint description of atmospheric~\cite{skatmos},
K2K~\cite{k2k}, solar~\cite{sksol,sno2} and reactor~\cite{kland,chooz}
data requires that all the three known neutrinos take part in the
oscillation process.  The mixing parameters are encoded in the $3
\times 3$ lepton mixing matrix which can be conveniently parametrized
in the standard form:
\begin{equation}
    U=\left(\begin{array}{ccc}
    1&0&0 \\ 
    0& {c_{23}} & {s_{23}} \\
    0& -{s_{23}}& {c_{23}}
    \end{array}\right)\times\left(
    \begin{array}{ccc} 
    {c_{13}} & 0 & {s_{13}}e^{i {\delta}}\\
    0&1&0\\ 
    -{ s_{13}}e^{-i {\delta}} & 0  & {c_{13}}
    \end{array}\right)
    \times \left(\begin{array}{ccc}
    c_{21} & {s_{12}}&0\\
    -{s_{12}}& {c_{12}}&0\\
    0&0&1\end{array}\right)
    \label{eq:evol.2} 
\end{equation}
where $c_{ij} \equiv \cos\theta_{ij}$ and $s_{ij} \equiv
\sin\theta_{ij}$. Note that the two Majorana phases are not included
in the expression above since they do not affect neutrino
oscillations. The angles $\theta_{ij}$ can be taken without loss of
generality to lie in the first quadrant, $\theta_{ij} \in [0,\pi/2]$.  

There are two possible mass orderings, which we denote as ``normal''
and ``inverted''.  In the normal scheme $m_1<m_2<m_3$ while in 
the inverted one $m_3< m_1<m_2$. 
The two orderings are often referred to in terms of
$\sgn(\Delta m^2_{31})$.

In total the three-neutrino oscillation analysis involves seven
parameters: 2 mass differences, 3 mixing angles, the CP phase and the
sign of $\Delta m^2_{31}$. Generic three-neutrino oscillation effects
include:
\begin{myitemize}
  \item coupled oscillations with two different oscillation lengths;
  \item CP violating effects;
  \item difference between Normal and Inverted schemes.
\end{myitemize}
The strength of these effects is controlled by the values of the ratio
of mass differences $\alpha \equiv \Delta m^2_{21}/|\Delta m_{31}^2|$, by the
mixing angle $\theta_{13}$ and by the CP phase $\delta$. 

For solar and atmospheric oscillations, the required mass differences
satisfy:
\begin{equation} \label{eq:deltahier}
    \Delta m^2_\odot = \Delta m^2_{21} \ll 
|\Delta m_{31}^2|=\Delta m^2_{\rm atm}.
\end{equation}
Under this condition, the joint three-neutrino analysis simplifies and
we have:
\begin{myitemize}
  \item for solar and KamLAND neutrinos, the oscillations with the
    atmospheric oscillation length are completely averaged and the
    survival probability takes the form:
    \begin{equation}
	P^{3\nu}_{ee}
	=\sin^4\theta_{13}+ \cos^4\theta_{13}P^{2\nu}_{ee} 
	\label{eq:p3}
    \end{equation}
    where in the Sun $P^{2\nu}_{ee}$ is obtained with the modified sun
    density $N_{e}\rightarrow \cos^2\theta_{13} N_e$. So the analyses
    of solar data constrain three of the seven parameters: $\Delta
    m^2_{21}, \theta_{12}$ and $\theta_{13}$. The effect of
    $\theta_{13}$ is to decrease the energy dependence of the solar
    survival probability;
  \item for atmospheric and K2K neutrinos, the solar wavelength is too
    long and the corresponding oscillating phase is negligible. As a
    consequence, the atmospheric data analysis restricts $\Delta
    m^2_{31}\simeq \Delta m^2_{32}$, $\theta_{23}$ and $\theta_{13}$,
    the latter being the only parameter common to both solar and
    atmospheric neutrino oscillations and which may potentially allow
    for some mutual influence. The effect of $\theta_{13}$ is to add a
    $\nu_\mu\rightarrow\nu_e$ contribution to the atmospheric
    oscillations;
  \item at CHOOZ~\cite{chooz} the solar wavelength is unobservable if
    $\Delta m^2< 8\times 10^{-4}~\eVq$ and the relevant survival
    probability oscillates with wavelength determined by $\Delta
    m^2_{31}$ and amplitude determined by $\theta_{13}$. 
\end{myitemize}

In this approximation, the CP phase is unobservable. In principle
there is a dependence on the Normal versus Inverted orderings due to
matter effects in the Earth for atmospheric neutrinos. However, this
effect is controlled by the mixing angle $\theta_{13}$, which is
constrained to be small by the combined analysis of CHOOZ reactor and
atmospheric analysis. As a consequence, this effect is too small to be
statistically meaningful in the present analysis. 

\begin{figure}[t] \centering
    \includegraphics[width=3.in]{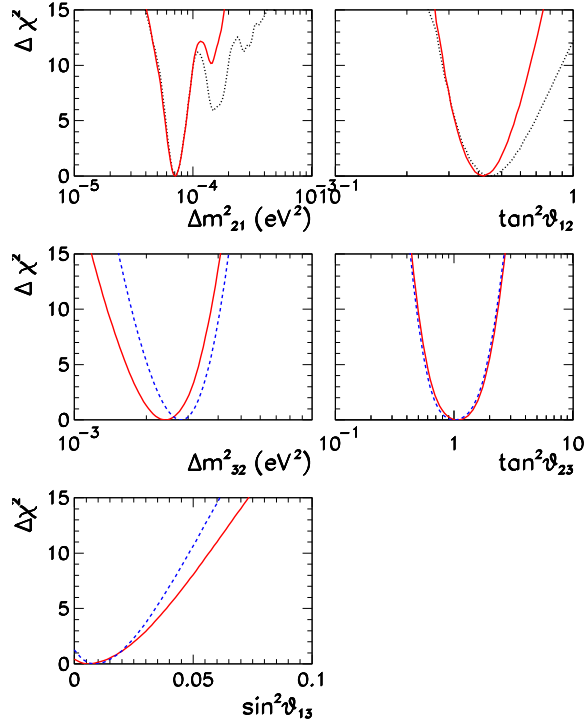}
    \caption{\label{fig:chiglo}%
      Global $3\nu$ oscillation analysis. Each panel on the left 
      shows the dependence of $\Delta\chi^2$ on each of the five
      parameters from the global analysis (full line) compared to the
      bound prior to the inclusion of the new ATM analysis (dashed
      blue line) and SNO data (dotted black line).}
\end{figure}

In Fig.~\ref{fig:chiglo} we plot the individual bounds on each of the
five parameters derived from the global analysis.  To illustrate the
impact of SNOII and the new ATM analysis we also show the
corresponding bounds when either SNOII and the new ATM fluxes are not
included in the analysis.  In each panel the displayed $\chi^2$ has
been marginalized with respect to the undisplayed parameters.

Quantitatively we find the following $3\sigma$ CL allowed ranges:
\begin{align} \label{eq:sranges}
    5.2 \leq \Delta m^2_{21}/{10^{-5}~\eVq} & \leq 9.8 \,, & \quad
    0.29 \leq \tan^2\theta_{12} & \leq 0.64 \,, \nonumber
    \\
    1.4 \leq \Delta m^2_{32}/{10^{-3}~\eVq} & \leq 3.4 \,, & \quad
    0.49 \leq \tan^2\theta_{23} & \leq 2.2 \,, \\
    && \sin^2\theta_{13} & \leq 0.054 \,. \nonumber
\end{align}
These results can be translated into our present knowledge of the
moduli of the mixing matrix $U$:
\begin{equation}
    |U| =\left(\begin{array}{ccc}
    0.78-0.88&0.47-0.62&<0.23 \\
    0.18-0.55&0.40-0.73&0.57-0.82\\ 
    0.19-0.55&0.41-0.75&0.55-0.82
\end{array}\right)
\end{equation}
which presents a structure
\begin{equation}
    |U| \simeq \left(\begin{array}{ccc}
    \frac{1}{\sqrt{2}}(1+{ \lambda}) &
    \frac{1}{\sqrt{2}}(1-{ \lambda}) &
    { \epsilon} \\
    \frac{1}{{2}}(1-{\lambda}+{\Delta} 
    +{ \epsilon}\,{\cos\delta}) &
    \frac{1}{{2}}(1+{\lambda}+{\Delta} 
    -{\epsilon}\,{\cos\delta}) &
    \frac{1}{\sqrt{2}}(1-{\Delta}) \\
    \frac{1}{{2}}(1-{\lambda}-{\Delta} 
    -{\epsilon}\,{\cos\delta}) &
    \frac{1}{{2}}(1+{\lambda}-{\Delta} 
    +{\epsilon}\,{\cos\delta}) &
    \frac{1}{\sqrt{2}}(1+{\Delta})
\end{array}\right)
\end{equation}
with $1\sigma$ ranges
\begin{equation}
    \lambda=0.23\pm0.03 \,, \quad
     \Delta=0\pm0.08 \,, \quad
    \epsilon\le 0.02 \,, \quad
    -1 \le \cos\delta\le 1 \,.
\end{equation}

\subsection{LSND and Sterile Neutrinos}

Together with the results from the solar and atmospheric neutrino 
experiments, we have one more piece of evidence pointing towards the
existence of neutrino masses and mixing: the LSND experiment, which
found evidence of $\overline{\nu}_\mu \rightarrow \overline{\nu}_e$
neutrino conversion with $\Delta m^2\geq 0.1~\eVq$. All these data can
be accommodated into a single neutrino oscillation framework only if
there are at least three different scales of neutrino mass-squared
differences. This requires the existence of a fourth light neutrino,
which must be {\it sterile} in order not to affect the invisible $Z^0$
decay width, precisely measured at LEP.

One of the most important issues in the context of four-neutrino
scenarios is the neutrino mass spectrum. There are six possible
four-neutrino schemes which can in principle accommodate the results
of solar and atmospheric neutrino experiments as well as the LSND
result. They can be divided in two classes: (3+1) and (2+2). In the
(3+1) schemes, there is a group of three close-by neutrino masses that
is separated from the fourth one by a gap of the order of 1~\eVq,
which is responsible for the SBL oscillations observed in the LSND
experiment. In (2+2) schemes, there are two pairs of close masses
separated by the LSND gap. The main difference between these two
classes is the following: if a (2+2)-spectrum is realized in nature,
the transition into the sterile neutrino is a solution of either the
solar or the atmospheric neutrino problem, or the sterile neutrino
takes part in both. This is not the case for a (3+1)-spectrum, where
the sterile neutrino could be only slightly mixed with the active ones
and mainly provide a description of the LSND result.

\begin{figure}[t] \centering
    \includegraphics[width=0.6\textwidth]{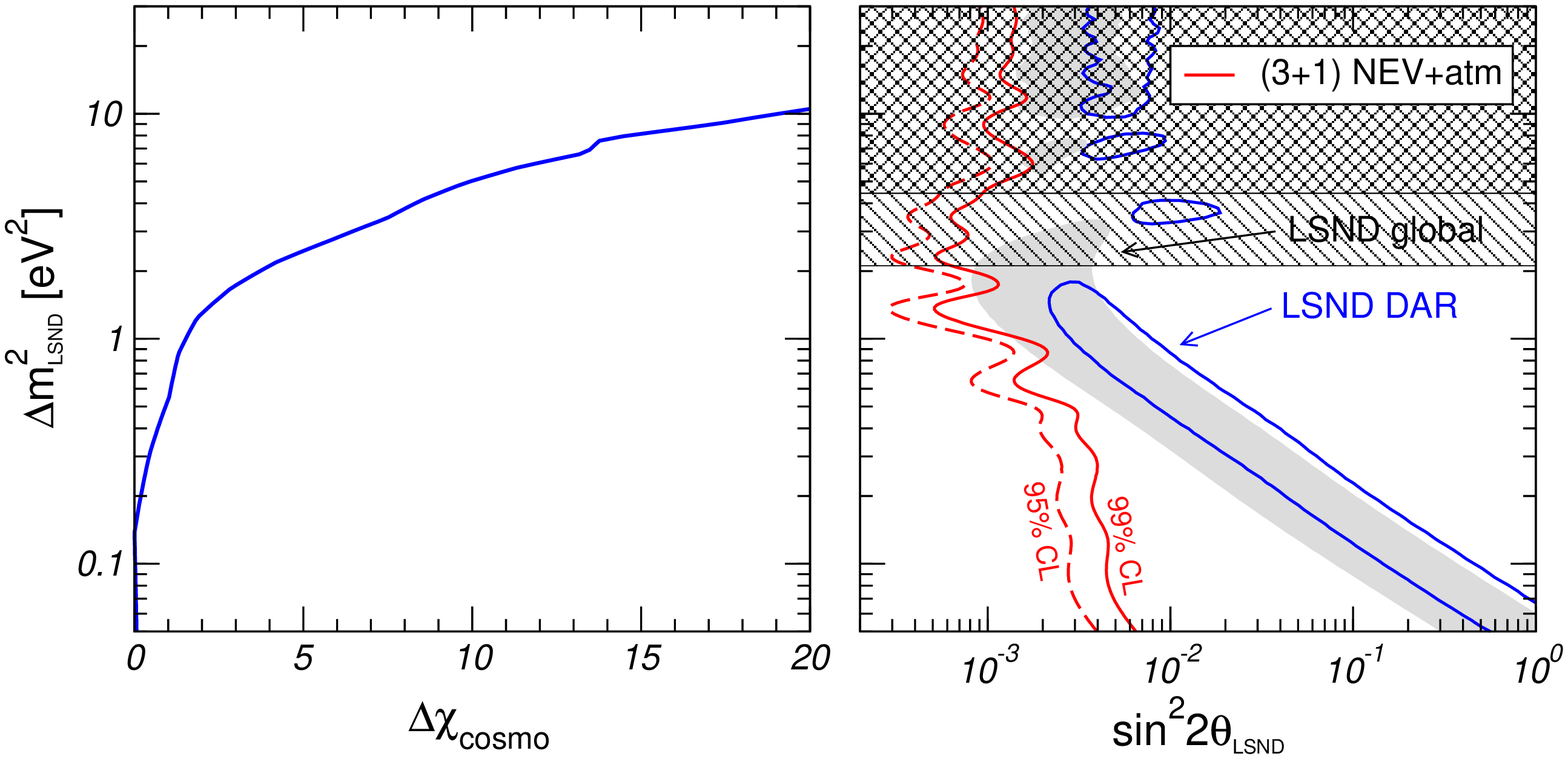}
    \includegraphics[width=0.35\textwidth]{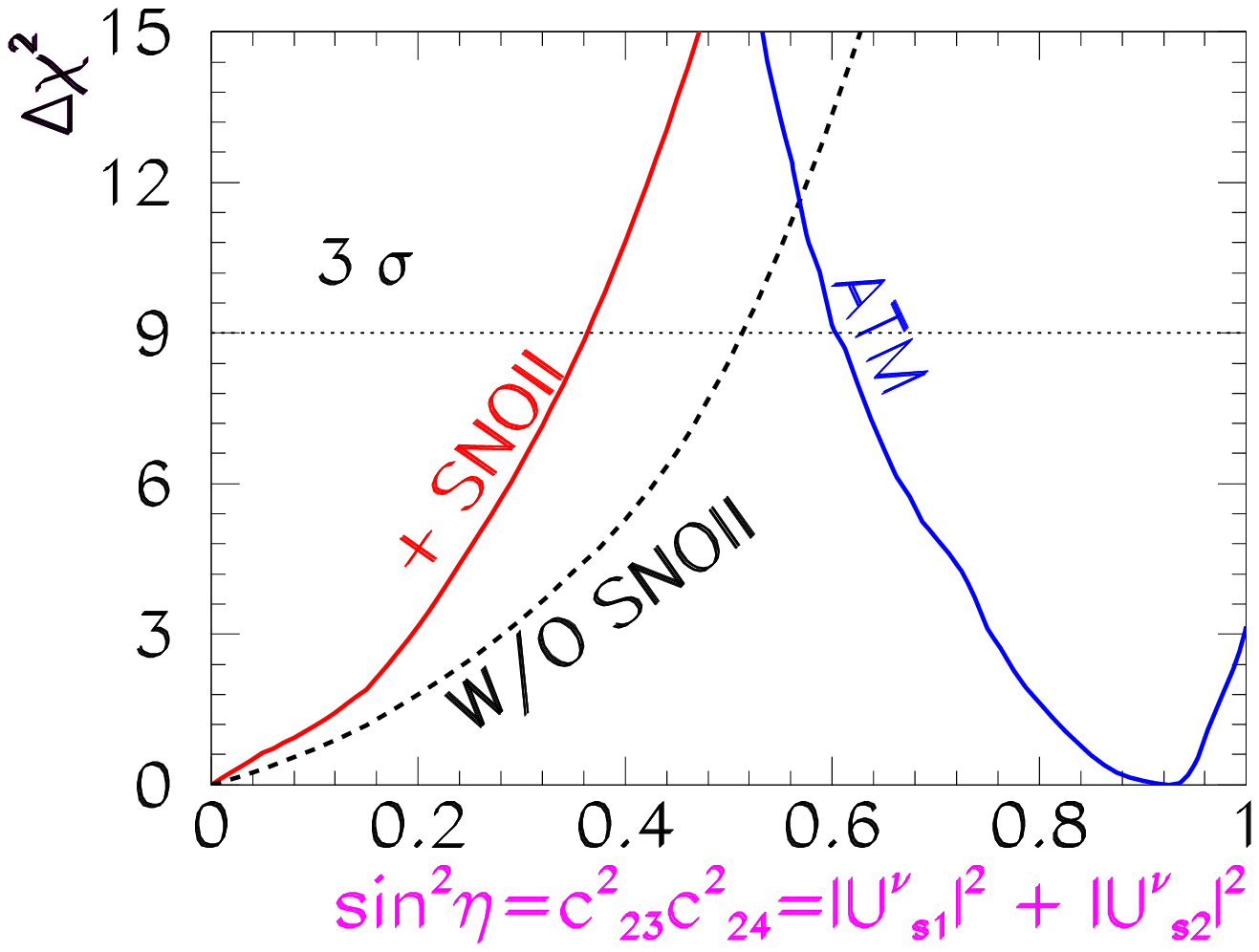}
    \caption{\label{fig:ster}%
      \textit{Left}: Status of the 3+1 oscillation scenarios.
      \textit{Right}: Present status of the bounds on the
      active-sterile admixture from solar and atmospheric neutrino
      data in (2+2)-models.}
\end{figure}

The phenomenological situation at present is that none of the
four-neutrino scenarios are favored by the data. Concerning
(2+2)-spectra, they are ruled out by the existing constraints from
the sterile oscillations in solar and atmospheric data. As for 
(3+1)-spectra, they are disfavored by the incompatibility between the
LSND signal and and the negative results found by other short-baseline
laboratory experiments. There is also a constraint on the possible
value of the heavier neutrino mass in this scenario from their
contribution to the energy density in the Universe which is presently
constrained by cosmic microwave background radiation and large scale
structure formation data~\cite{cosmo}.

We show in Fig.~\ref{fig:ster} the latest results of the analysis of
neutrino data in these scenarios. In the left and central panel we
summarize the results from Ref.~\cite{michele} on the (3+1)
scenarios; we see that after the inclusion of the cosmological bound
there is only a marginal overlap at 95\% CL between the allowed LSND
region and the excluded region from SBL+ATM experiments. The right
panel illustrates the status of the (2+2) scenarios. At present, the
lower bound on the sterile component from the analysis of atmospheric
data and the upper bound from the analysis of solar data do not
overlap at more than $4\sigma$. The figure also illustrates the effect
of the inclusion of the SNOII in this conclusion. 

\begin{figure}[t] \centering
    \includegraphics[width=3in]{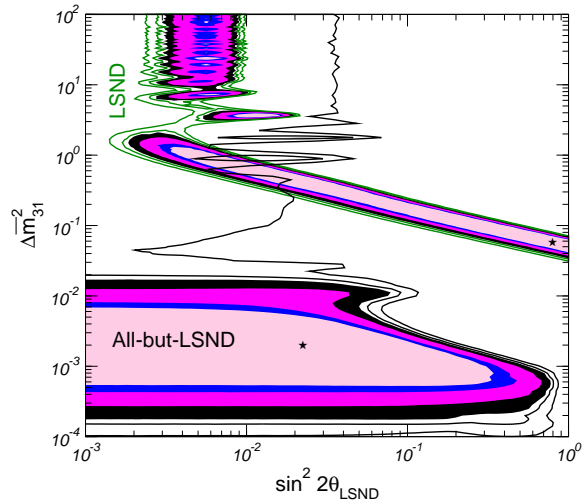}
    \caption{\label{fig:cptstatus}%
      90\%, 95\%, 99\%, and $3\sigma$ CL allowed regions (filled) in
      required to explain the LSND signal together with the
      corresponding allowed regions from our global analysis of
      all-but-LSND data. The contour lines correspond to $\Delta\chi^2
      = 13$ and 16 (3.2$\sigma$ and 3.6$\sigma$, respectively).}
\end{figure}

Alternative explanations to the LSND result include the possibility of
CPT violation~\cite{CPT}, which implies that the masses and mixing
angles of neutrinos may be different from those of antineutrinos. We
have performed an analysis of the existing data from solar,
atmospheric, long baseline, reactor and short baseline data in the
framework of CPT violating oscillations~\cite{ourcpt}. The summary of
the results of this analysis is presented in Fig.~\ref{fig:cptstatus},
which shows clearly that there is no overlap below the $3\sigma$ level
between the LSND and the all-but-LSND allowed regions. We also note
that that the all-but-LSND region is restricted to $\Delta
\overline{m}_{31}^2 = \Delta m^2_{\rm LSND}<0.02~\eVq$, whereas for
LSND we always have $\Delta \overline{m}_{31}^2 = \Delta m^2_{\rm
LSND} > 0.02~\eVq$.

\vspace{-3mm}
\section*{Acknowledgments}

This work was supported in part by the National Science Foundation
grant PHY0098527.  MCG-G is also supported by Spanish Grants No.\
FPA-2001-3031 and CTIDIB/2002/24.

\end{document}